\documentclass{article}
\usepackage[utf8]{inputenc}
\usepackage{graphics,graphicx,amssymb,amsmath,dcolumn,bm,url,amsbsy,pgf}
\usepackage[none]{hyphenat}
\usepackage{microtype}
\usepackage{setspace}
\usepackage{geometry}
\usepackage{pslatex} 
\usepackage{indentfirst}
\usepackage{array}
\usepackage{booktabs}
\usepackage{authblk}
\providecommand{\keywords}[1]{\textbf{{Keywords:}} #1}

\title{Isochronous bifurcations in a two-parameter twist map}
\author[1]{Michele Mugnaine \thanks{mmugnaine@gmail.com}}
\author[1]{Bruno B. Leal}
\author[2]{Alfredo M. Ozorio de Almeida}
\author[3]{Ricardo L. Viana}
\author[1]{Iber\^e L. Caldas}%
\affil[1]{Institute of Physics, University of S\~ao Paulo, S\~ao Paulo, SP, 05508-090, Brazil}
\affil[2]{Brazilian Center for Research in Physics, Rio de Janeiro, RJ, 22290-180, Brazil}
\affil[3]{Department of Physics, Federal University of Paran\'a, Curitiba, PR, 81531-980, Brazil}
\date{}

\begin{document}
	
	\maketitle

 \begin{abstract}
Isochronous islands in phase space emerge in twist Hamiltonian systems as a response to multiple resonant perturbations. According to the Poincaré-Birkhoff theorem, the number of islands depends on the  system characteristics and the perturbation. We analyze, for the two-parameter standard map, also called two-harmonic standard map, how the island chains are modified as the perturbation amplitude increases. We identified three routes for the transition from one chain, associated with one harmonic, to the chain associated with the other harmonic, based on a combination of pitchfork and saddle-node bifurcations. These routes can present intermediate island chains configurations. Otherwise, the destruction of the islands always occurs through the pitchfork bifurcation. 
 \end{abstract}
\section{Introduction}

Non-integrable Hamiltonian systems appear in a vast number of physical applications, and their general analysis is usually difficult due to the large dimensionality of the corresponding phase space. Hence, numerical investigations often use low-dimensional systems in discrete time, due to their inherent simplicity and fast computational simulations. Among these, it stands out the so-called standard (or Chirikov-Taylor) map \cite{chirikov1979}, which is by itself a simplified model of various physical problems, like particle dynamics in accelerators \cite{izraelev1980}, charged particle confinement in mirror magnetic traps \cite{chirikov1979}, microwave ionization of Rydberg atoms \cite{benvenuto1994}, just to mention some representative examples. Moreover, a wide class of dynamical systems and maps can be locally reduced to the standard map \cite{lichtenberg}. Finally, it is worth mentioning that the standard map has been used to prove many key results in the theory of quasi-integrable dynamical systems, like the transition to global stochasticity \cite{greene1979}.

For twist Hamiltonian systems, multiple resonant perturbation with the same frequency generate sets of distinct chain of islands, in the same region of phase space. These islands are called isochronous and they present the same winding number, i.e., they have the same average speed around an invariant circle \cite{ketoja1989}. The number of island chains varies as a function of the wave parameters and the amplitude of the resonant terms. According to the Poincaré-Birkhoff fixed point theorem, any resonance with a winding number equal to $r/s$ produces $2k$ periodic orbits with period $s$, where $k$ is a positive integer number\cite{lichtenberg}. Half of these orbits is stable and half unstable. When $k > 1$, we have isochronous islands. The set of isochronous islands surrounding one stable orbit form an isochronous chain. In this sense, $k$ indicates the number of chains and $r$ represents the number of islands in each chain. Although the Poincaré-Birkhoff fixed point theorem does not determine the value of $k$, we generally find in the literature systems that present just one chain	with $r$ islands. However, for some systems, more than one chain has already been observed (see, for example, references \cite{walker1969,van1988}).

Isochronous island chains have been recognized in several areas of physics, such as	nonlinear oscillators \cite{walker1969}, molecular physics  \cite{egydio1992}, electron beam interactions with electrostatic waves \cite{sousa2013}, Plasma Physics \cite{wu2019,evans2021PRL,leal2023}, and periodic lattices \cite{lazarotto2022}. For the case of wave-particle interactions, isochronous island chains were reported in a nearly integrable model \cite{sousa2013} describing the dynamics of relativistic charged	particles moving in a uniform magnetic field and kicked by standing electrostatic pulses. Different island chains were identified for perturbation terms with the same winding number that generate isochronous islands in the same region of phase space.

Recently, for plasma confined in tokamaks, magnetic field line configurations have been investigated with natural and external modes with different wave numbers and the same winding number, thus resonant at the same rational magnetic surface. As the amplitude of an externally applied magnetic field perturbation is increased, the topology of resonant helical magnetic islands are altered through what is called heteroclinic bifurcations \cite{evans2021PRL,leal2023}. The heteroclinic bifurcation and the emergence of new islands are a plasma response to non-axisymmetric perturbation \cite{wu2019,evans2021PRL,evans2021}. As concluded by Wu and co-workers \cite{wu2019}, the reported bifurcations in tokamaks are interesting, but a dynamic explanation was not available yet. In addition to these bifurcations having been observed in solutions of differential equations, they have also been identified in symplectic maps \cite{leal2023}.

With the aim of proposing a two-parameter map that can have isochronous islands and bifurcations that change the stability of fixed points, we propose the generalization of the extended standard map (ESM) with the idea of providing a simple map that represents the competition and coupling between two different modes. The extended standard map can be obtained from the Frenkel-Kontorova model when the critical behavior with a second harmonic of the external potential is analyzed \cite{greene1987} or with the inclusion of a second harmonic with half of the spatial period in the standard map (SM) \cite{greene1990}. The majority of analysis of the ESM is about the breakup of the invariant tori, especially the last irrational one with a golden mean winding number. The main results about these analyses can be found in References \cite{greene1990,ketoja1989,baesens1994}. Differently, in this survey, we restrict our analysis to the resonant tori with a rational winding number, i.e., the islands in the phase space.

From our results, we show that the generalized version of ESM allows the study of competition between two arbitrary resonant modes. As stated by the Poincaré-Birkhoff theorem, for a perturbed system, there are an even number of fixed points, where half of them are stable, and the other is unstable \cite{lichtenberg}. The stable points, named elliptic points, are surrounded by resonant islands which are stable periodic orbits in the phase space \cite{sousa2015}. In this manuscript, we defined the number of chains of elliptic points as the mode of the system. Therefore, based on the competition of resonant modes, we propose the generalized ESM as a model to reproduce isochronous bifurcations, i.e., the emergence of isochronous islands by bifurcations, as well as the bifurcations observed in the Poincaré section of field line integration codes. From the phase space analysis and from counting the number of elliptic points, we can identify the transitions from one mode to another, as well as the existence of possible intermediate modes. The observed transitions due to the isochronous bifurcations are consequences of saddle-node and pitchfork bifurcations.


This paper is organized as follows: The generalized extended standard map also called the two-harmonic standard map, is presented in Section 2. The analysis of isochronous bifurcations is presented in Section 3 as well as the routes to emergence and destruction of stable orbits. Our Conclusions are in the last section. 

\section{Two-harmonic standard map}
	
The extended form of the standard map can be obtained by the consideration of a one-dimensional lattice of particles, where each one interacts elastically with the nearest neighbor \cite{greene1987,johannesson1988}. In the equilibrium and for a lattice periodicity equal to one, Greene and coauthors obtained \cite{greene1987},
	\begin{eqnarray}
		\begin{aligned}
			y_{n+1}&=y_n-\dfrac{K_1}{2\pi} \sin(2 \pi x_n)-\dfrac{K_2}{4 \pi} \sin (4\pi x_n),\\
			x_{n+1}&=x_n+y_{n+1},
			\label{map1}
		\end{aligned}
	\end{eqnarray}
an area-preserving twist map which would be called extended standard map \cite{greene1990,ketoja1989,ketoja1990} or two-harmonic twist map \cite{baesens1994,lomeli2006}. The extended map was widely investigated and many of its properties are already known and well established. Greene and Mao, Ketoja and MacKay discussed, in two different papers, the breakup of the golden torus in the extended map and obtain its critical line in the parameter space \cite{greene1990,ketoja1989}. Beasen and MacKay continued to study and investigate the transition of sequences of periodic orbits \cite{baesens1994}, while Ketoja analyzed the breakup of tori with arbitrary frequency \cite{ketoja1990}. More recently, transport analysis and statistical characterization were applied to the extended map \cite{lomeli2006,cetin2022}.
	
In this paper, we introduce a map that describes the interaction between two arbitrary harmonic modes. Unto that, we generalized the map (\ref{map1}) and the two-dimensional maps becomes,
\begin{eqnarray}
	\begin{aligned}
		y_{n+1}=&y_n-\dfrac{K_1}{2\pi~ m_1} \sin(2 \pi~ m_1 x_n) -\dfrac{K_2}{2 \pi~ m_2}\sin(2\pi~ m_2 x_n),\\
		x_{n+1}=&x_n+y_{n+1}
		\label{map2}
	\end{aligned}
\end{eqnarray}
where $K_1$ and $K_2$ are the amplitudes of the modes $m_1$ and $m_2$, respectively. Both variables are taken mod 1, and we establish $x\in[-0.5,0.5]$ and $y\in[0,1]$.  In this paper, we consider $K_1, K_2 \in\mathbb{R}$ and $m_1,m_2 \in \mathbb{Z^*+}$. Since for $K_2=0$ and $m_1=1$ we recover the Chirikov-Taylor map, we named the map (\ref{map2}) as two-harmonic standard map. 
	
\begin{figure}[!h]
	\begin{center}
		\includegraphics[width=0.7\textwidth]{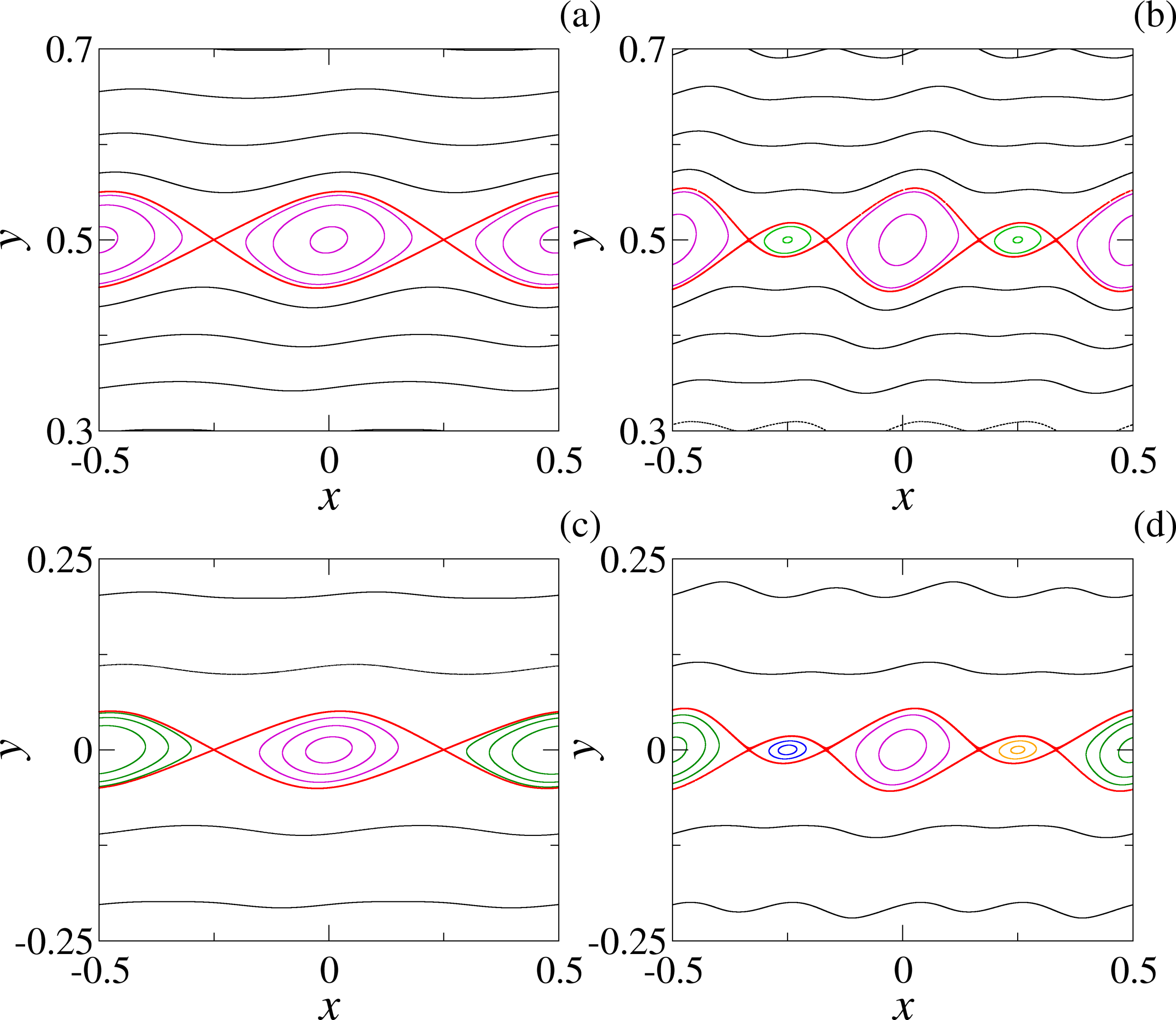}		
		\caption{Phase spaces for the two-harmonic standard map with $m_1=2$, $m_2=4$, $K_1=0.1$ and two configurations with different values of $K_2$. The phase space in (a), with $K_2=0.02$, represents the case where the first mode is predominant with two non-isochronous islands. In this case, we observe one chain of islands with period 2. For (b), we have the predominance of the second mode, $K_2=0.2$ and we observe two chains of isochronous islands. Two chains of isochronous islands can also be observed in (c), where two islands of period 1 are present, for $K_2=0.02$. If we set $K_2=0.2$, we have the phase space in (d), where four isochronous islands of period 1 are in $y=0$. The apparent separatrices (thicker red lines) for all phase spaces are actually thin chaotic regions. } 
		\label{fig1}
	\end{center}
\end{figure}

Our main goal with this survey is to simulate the competition between two resonant modes. For this, we consider the term defined by $K_1$ and $m_1$ as the first mode while the second mode, described with the parameters $K_2$ and $m_2$ is the new mode that is competing to become the predominant one. The competition of modes only exists if the resonances present the same winding number. As an example, we set $m_1=2$, $m_2=4$, $K_1=0.1$ and choose two values of $K_2$ indicating two different amplitudes for the second mode. To differentiate between isochronous and non-isochronous islands, we present resonant scenarios in different regions of the phase space. First, we observe the islands in $y=0.5$. These islands present winding numbers $\omega=1/2$ and period 2. The other studied region is $y=0$, where the islands in this line have winding numbers and periods equal to unity. The phase spaces are shown in Figure \ref{fig1}
	
For the first phase space in Figure \ref{fig1} (a), $K_1=0.1$ and $K_2=0.02$. In this configuration, the first mode is predominant and, since $\omega=0.5$, we observe a chain of islands of period 2. Since the islands are non-isochronous, one initial condition in one of the islands will visit both islands. With the increase of the amplitude of the second mode, we have the scenario presented in Figure \ref{fig1} (b). In this case, $K_2=0.2$ and we can observe that other two islands emerge. As a result, we observe two chains of islands of period 2. The pink islands, centered in $(x,y)=(0,0)$, $(x,y)=(0.5,0)$, are isochronous related to the green centered in $(x,y)=(\pm 0.25,0)$. From this scenario, we observe the transition from  the mode $m_1=2$ with two islands for a configuration of four islands representing the mode $m_2=4$. But, since we are studying the line $y=0.5$, the islands have period 2, and as long as we observe four islands in Figure \ref{fig1} (b), we have two chains of elliptic points of period 2.

A visually similar but completely different scenario is presented in Figures \ref{fig1} (c) and (d). Differently from the scenario shown in Figure \ref{fig1} (a), we observe two isochronous islands in Figure \ref{fig1} (c), i.e., two islands of period 1. These islands present the same period 1 but are not connected, as we highlight with different colors for different islands. Increasing the amplitude of the second mode, we have the scenario presented in Figure \ref{fig1} (d). In this case, we also observe the emergence of the 4-mode configurations with four islands, but they are of period one. The new two islands have their centers in $x=\pm 0.25$ and we present them with blue and orange colors. Here, we have four isochronous islands of period 1, i.e., four different elliptic points. As for Figures \ref{fig1} (a) and (b), $K_2=0.02$ and $K_2=0.2$ for Figures \ref{fig1} (c) and (d), respectively.

\begin{figure}[!h]
	\begin{center}
		\includegraphics[width=1.0\textwidth]{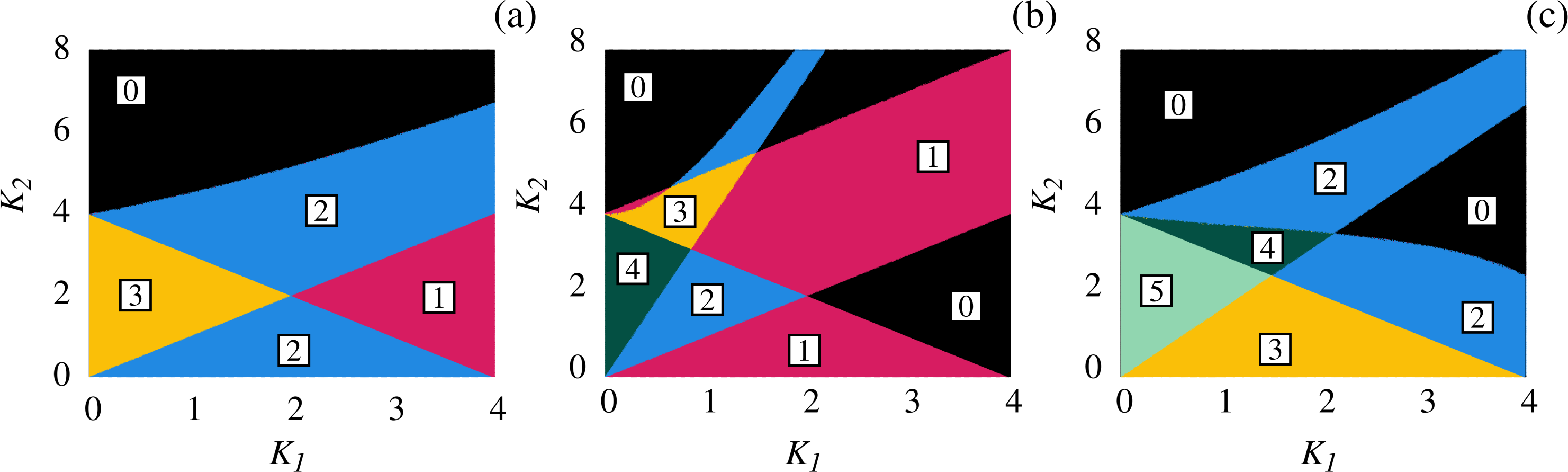}		
		\caption{Parameter space $K_1 \times K_2$ for the quantity of elliptic points in $y=0$ for (a) $(m_1,m_2)=(2,3)$, (b) $(m_1,m_2)=(1,4)$, (c) $(m_1,m_2)=(3,5)$. The colors black, pink, blue, orange, dark green and light-green indicate the quantity of 0, 1, 2, 3, 4 and 5 elliptic points, respectively.}
		\label{fig2}
	\end{center}
\end{figure}

From Figure \ref{fig1}, we conclude there is a transition between the two modes represented by the two harmonics of the map (\ref{map2}). It should be a critical value where the second mode emerges and becomes prevalent in the phase space. To analyze the transitions from mode $m_1$ to mode $m_2$, we calculate the number of different elliptic points at $y=0$ for different values of $K_{1,2}$ and $m_{1,2}$. In figure \ref{fig2} we present three parameter spaces $K_1 \times K_2$ for three different combinations of $m_1$ and $m_2$: (a) $m_1=2$ and $m_2=3$, (b) $m_1=1$ and $m_2=4$, and (c) $m_1=3$ and $m_2=5$. Each color represents the number of elliptic points indicated by the number in the box in each colored region.
	
Each parameter space in Figure \ref{fig2} shows different structures and arrangements of the colored regions. In Figure \ref{fig2} (a),  we present the parameter space for the competition between modes $m_1=2$ and $m_2=3$, which we will indicate as $2\to3$. We observe this transition for a specific region in the parameter space indicated by the boundary between blue and orange regions when the value of $K_2$ is increasing. However, we can observe other transitions, such as from mode 2 to 1, from 1 to 2, from 3 to 2, and from 2 to 0. Each transition is indicated by the boundary between the two respective colored regions. For $K_1>2$, we conclude the transition $2\to 3$ is not possible, instead, we observe the transition $2 \to 1$ and $1 \to 2$. The destruction of the islands is the same for every value of $K_1$: the two reminiscent islands are destroyed simultaneously.
	
From Figure \ref{fig2} (b), we observe the possibilities of chains when modes 1 and 4 are competing. It is a more complex scenario if we compare it to the parameter space of Figure \ref{fig2} (a). As a first big difference, we observe the transition $1 \to 4$ is not direct, there is the intermediate mode 2. Thus, the transition is $1 \to 2 \to 4$. However, as presented in the first parameter space, the transition between the two competing modes does not happen for any value of $K_1$ and $K_2$, it happens also only if $K_1<2$. We observe others transitions, as $4\to 3$, $1 \to 0$, $3\to 2$ and so on. Another difference, the destruction of the islands is not the same in the entire space. We observe two transitions: $1 \to 0$ and $2 \to 0$.
	
Lastly, analyzing the parameter space in Figure \ref{fig2} (c), we observe a directed transition between the competing modes 3 and 5. As in the parameter space in Figure \ref{fig2} (a), there is only one route to the destruction of the islands, the transition $2\to 0$. However, differently for the other cases, the mode $m_1=3$ does not return to be prevalent once it loses to the second mode, i.e., starting from a different mode and increasing the value of $K_2$, we do not observe the emergence of three islands of period one in the phase space. The parameter spaces for all combinations of $m_1$ and $m_2$, with $m_1\in[1,5]$ and $m_2\in[m_1+1,6]$ can be checked in the Supplementary material.	
	
Comparing all the parameter spaces, there is only one common aspect between them: the boundary defined by the line $K_2=4-K_1$. In order to comprehend this line, we can perform analytical analysis of the bifurcation of fixed points at the line $y=0$. The fixed point $(x^*,y^*)$ occurs for $y=0$ and for a value of $x$ which $(K_1/m_1) \sin(2 \pi~m_1 x)+(K_2/m_2) \sin(2 \pi ~m_2 x)=0$, where we obtain the trivial solution $(x^*,y^*)=(0,0)$. The computation of other fixed points depends on the values of $K_1$ and $K_2$ as the parity and the values of $m_1$ and $m_2$.
	
The type of the bifurcation that happens at each fixed point is obtained by the computation of the eigenvalues $\lambda$ of the Jacobian matrix of the map (\ref{map2}) calculated in $(x^*,y^*)$. From this, we obtain,
	
\begin{eqnarray}
	\lambda= \dfrac{1}{2}\left\{2-K_1 \cos(2 \pi m_1 x_n) - K_2\cos(2 \pi m_2 x_n) \pm \sqrt{\left[K_1 \cos(2 \pi m_1 x_n) + K_2\cos(2 \pi m_2 x_n)-2\right]^2-4}\right\}
	\label{lambda}
\end{eqnarray}
	
From Equation (\ref{lambda}), we conclude that the point $(x^*,y^*)=(0,0)$ is elliptic ($\lambda \in \mathbb{C}$) and then becomes hyperbolic ($\lambda \in \mathbb{R}$) when $K_1+K_2=4$. The point $(x^*,y^*)=(0,0)$ is common for all combinations of $m_1$ and $m_2$. Thus, the line $K_2=4-K_1$ is present in all parameter spaces as the ones shown in Figure \ref{fig2}. 
	
\section{Isochronous bifurcations}
In the previous section, we study the bifurcation of fixed points analytically. We obtained few results which can not explain completely the bifurcation routes in the system. Now, we analyze the bifurcation of the fixed points and the change in islands topology numerically, investigating the phase space. We analyze the phase spaces for every combination of $m_1\in[1,5]$ and $m_2 \in[m_1+1,6]$ and we have obtained three different routes for the transition $m_1 \to m_2$ which we describe below. In all phase spaces, the red thicker lines indicate chaotic regions. Since the parameters values are small, the chaotic regions are very thin and resemble the shape of a separatrix curve. For this reason, we emphasize the non-integrable characteristic of the system by highlighting the chaotic trajectories in the phase space by a thicker line.
	
\subsection{Route 1: pitchfork bifurcation}
	
\begin{figure}[!h]
	\begin{center}
		\includegraphics[width=.85\textwidth]{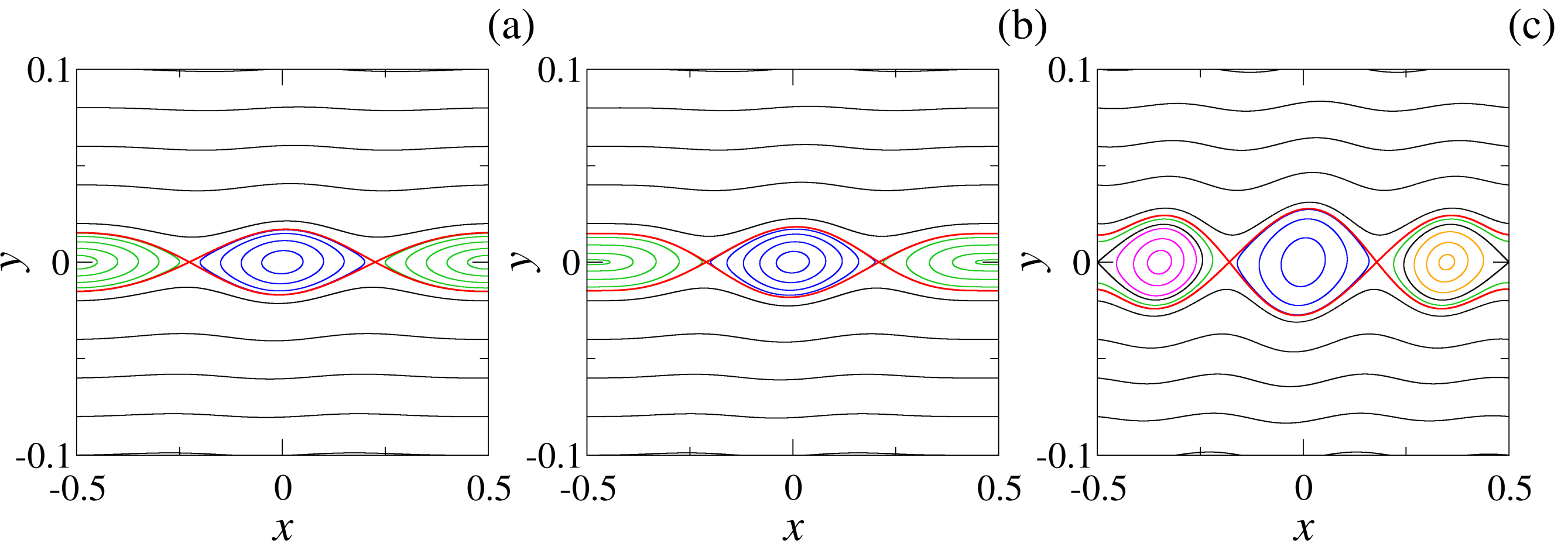}		
		\caption{Phase spaces for the transition $2 \to 3$. For all spaces, $m_1=2$, $m_2=3$, $K_1=0.01$ and (a) $K_2=0.005$, (b) $K_2=0.01$, (c) $K_2=0.05$. In (a) we have the predominance of the first mode, $m_1=2$, while in (b) we are in the imminence of the pitchfork bifurcation. For the phase space in (c), after the bifurcation, we observe the prevalence of the second mode $m_2=3$. The thicker line separatrices are thin chaotic layers. }
		\label{fig3}
	\end{center}
\end{figure}

The first route represents a scenario where the transition from mode $m_1$ to mode $m_2$ occurs with no intermediate modes and by a pitchfork bifurcation. Therefore, a certain number of fixed points change their stability and twice as many fixed points emerge with the opposite stability. This route can be observed in the directed transition $2\to 3$, present in the parameter space of Figure \ref{fig2} (a), as we can observe in the phase spaces of Figure \ref{fig3}.
	
From the sequence of phase spaces shown in Figure \ref{fig3}, we observe the elliptic point at $(x,y)=(-0.5,0)$ passes through a pitchfork bifurcation, becoming hyperbolic. Along with the change of stability, two elliptic points emerge and the mode $m_2=3$ becomes predominant. The same route is observed in Figure \ref{fig1} where the two hyperbolic points at $x=\pm0.25$ are replaced by two elliptic points and two hyperbolic points emerge aside. We also observe this route for the transitions $1\to2$, $2\to4$, $3\to4$, $3\to6$, $4\to5$, $4\to 6$ and $5 \to 6$. 
	
\subsection{Route 2: saddle-node/tangent bifurcation}
	
\begin{figure}[!h]
	\begin{center}
		\includegraphics[width=.85\textwidth]{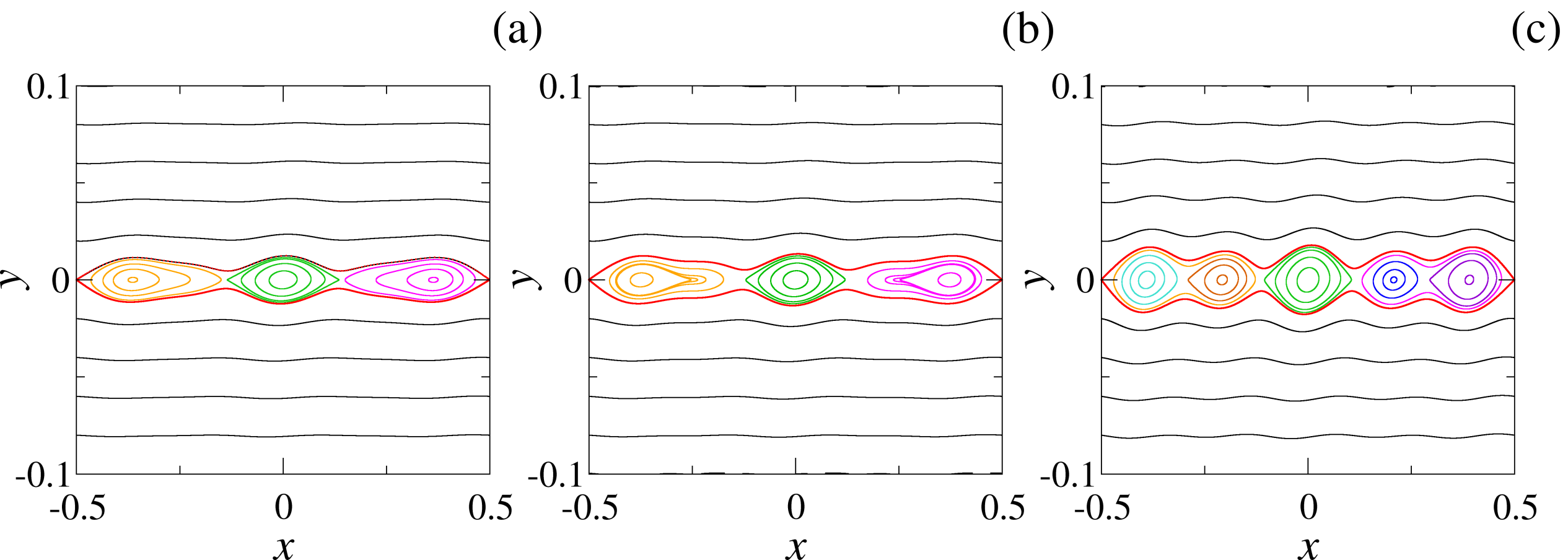}		
		\caption{Phases spaces for the transition $3 \to 5$. For (a) $K_2=0.01$ we observe three islands, representing the mode $m=3$. When $K_2=0.016$, the bifurcation is about to happen, as we can observe in (b). After the  emergence of two elliptic points and, consequently, two hyperbolic points, the width of the islands increases, and we have the scenario shown in (c), for $K_2=0.05$. For all phase spaces $K_1=0.01$ and the thicker lines are very thin chaotic layers.}
		\label{fig4}
	\end{center}
\end{figure}

The second observable route is composed only by a saddle-node/tangent bifurcation. As Route 1, there is no intermediate mode and the change from mode $m_1$ to mode $m_2$ occurs directly by the creation of elliptic and hyperbolic points by the same amount (saddle-node bifurcation).  We illustrate this route by the transition from mode $m=3$ to $m=5$, as shown by the sequence of phase spaces in Figure \ref{fig4}.
	
For $K_1=0.01$ and $K_2=0.01$, $m_1=3$ and $m_2=5$, we have the phase space in Figure \ref{fig4} (a) where we observe three islands. If the parameter $K_2$ slightly increases, we have the phase space in Figure \ref{fig4} (b), where the islands centered in $x\approx \pm 0.04$ are deformed, and the bifurcation is about to happen. The bifurcation in this scenario is the emergence of elliptic and hyperbolic points. In this case, two elliptic points and two hyperbolic points emerge simultaneously in the phase space, leading to a direct transition from mode 3 to 5. In Figure \ref{fig4} (c), for $K_2=0.05$, we observe 5 distinct islands related to the predominant mode $m_2=5$. This route was also identified for the transitions $1 \to 3$, $1 \to 5$ and $2 \to 6$.
	
\subsection{Route 3: intermediate mode with saddle-node and pitchfork bifurcation}			
	
\begin{figure}[!h]
	\begin{center}
		\includegraphics[width=.85\textwidth]{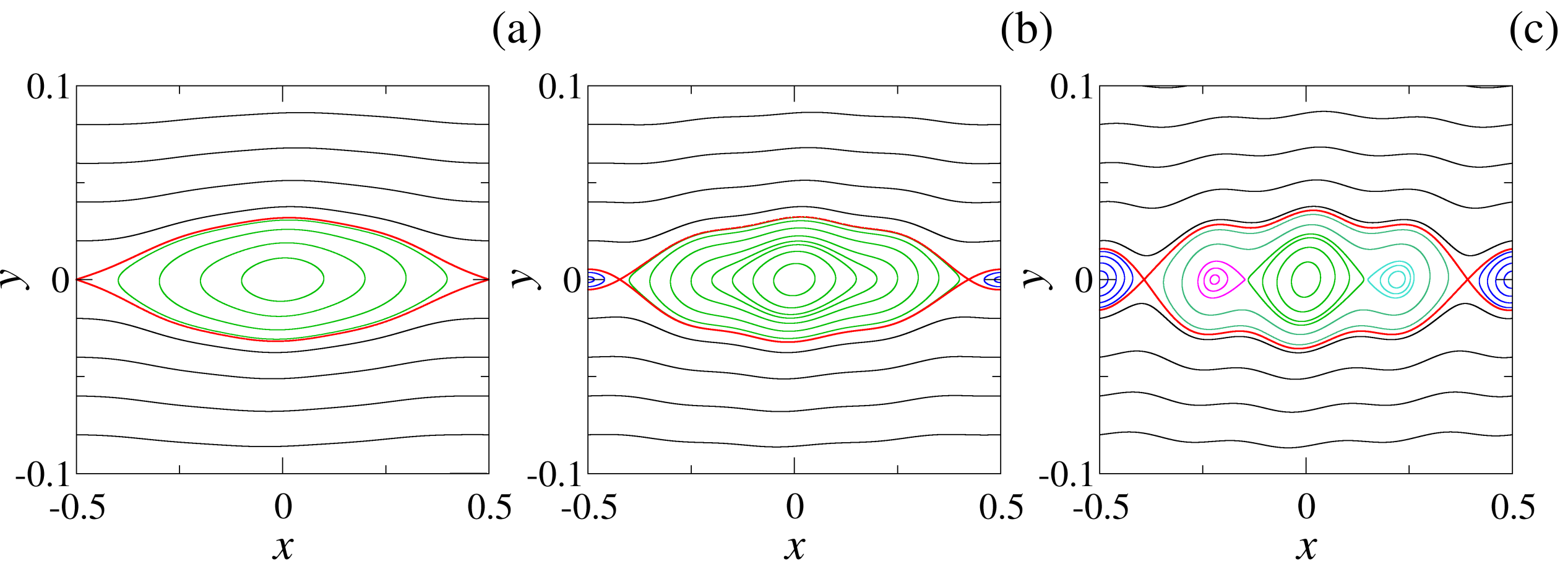}		
		\caption{Transition from $m_1=1$ to $m_2=4$ with intermediate mode. The amplitude of the first mode is, for all phase spaces, $K_1=0.01$ and it is predominant when $K_2=0.005$ in (a). The intermediate mode is visible when (b) $K_2=0.02$, where two isochronous islands are present in the phase space. For (c) $K_2=0.06$, the second mode is predominant and we observe four islands. The thicker red curve indicates the chaotic layer in the phase space.}
		\label{fig5}
	\end{center}
\end{figure}

Lastly, we have the route related to the parameter space in Figure \ref{fig2} (b) where there is an intermediate mode in the transition. Analyzing Figure \ref{fig2} (b), we choose $K_1=0.01$, $m_1=1$, $m_2=4$ and three different values of $K_2$ to exemplify the route with intermediate mode. The phase spaces related to the Route 3 are shown in Figure \ref{fig5}.
	
For the first phase space in Figure \ref{fig5} (a) we chose $K_2=0.005$ and we observe one island, related to the predominant mode $m_1=1$. When $K_2=0.02$, we observe two islands centered in $x=0$ and in $x=0.5$, as we see in Figure \ref{fig5} (b). The first bifurcation related to this intermediate transition is a pitchfork bifurcation, where the hyperbolic point at $x=0.5$ becomes elliptic and two other hyperbolic points emerge. The transition $2 \to 4$ occurs following two saddle-node bifurcations, as we can observe in Figure \ref{fig5} (c) for $K_2=0.06$. In this scenario, inside the big island centered in $x=0$ of Figure \ref{fig5} (b), two other islands emerge, such as two hyperbolic points, resulting in four elliptic points. This route was also observed for the transitions $1 \to 6$ and $2 \to 5$.

\subsection{Destruction route of the islands}	
	
\begin{figure}[!h]
	\begin{center}
		\includegraphics[width=.85\textwidth]{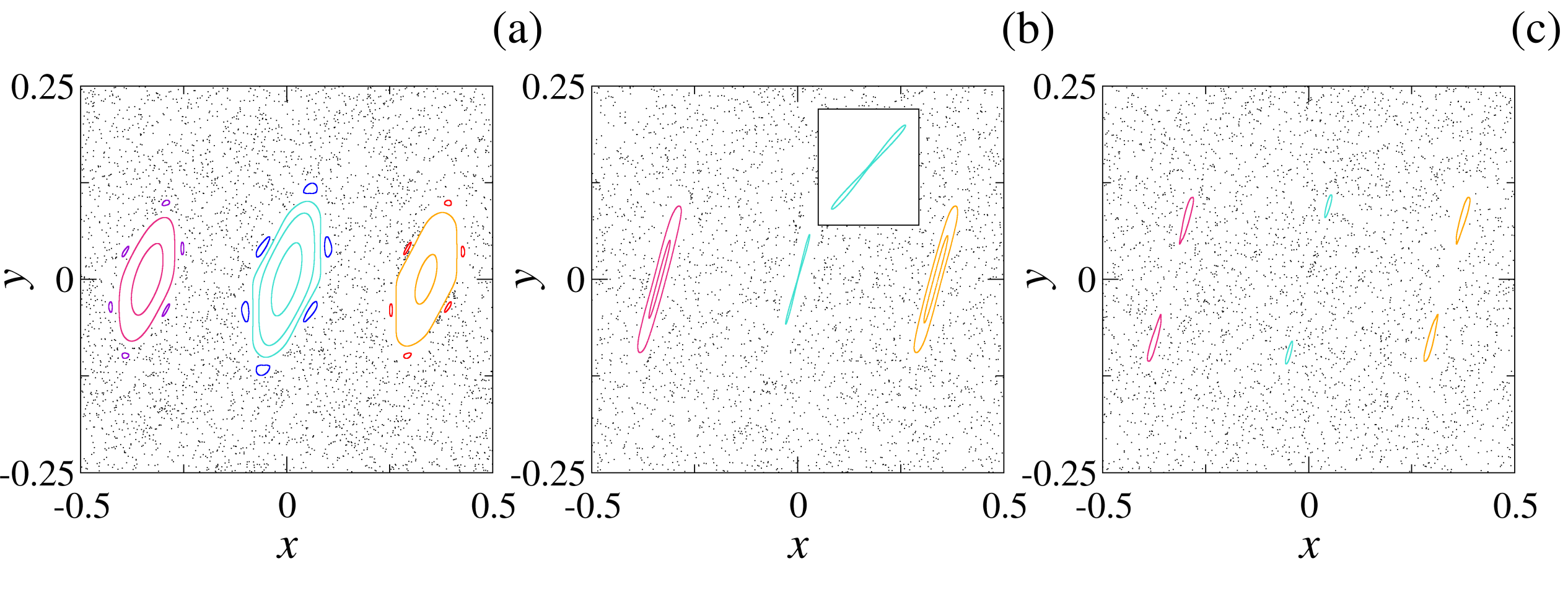}		
		\caption {Destruction route for the islands of the Standard Map with two modes. The islands become elongated in the $y$-direction as $K_2$ increases, as we observe in (a) for $K_2=1.5$. Eventually, the elliptic points in the center of islands go through a pitchfork bifurcation and become hyperbolic points, as shown in (b) for $K_2=4.0$ and the inset magnification. After the bifurcation, two islands emerge in the place, as shown in (c) for $K_2=4.5$. For all phase spaces $m_1=2$, $m_2=3$ and $K_1=0.1$.}
		\label{fig6}
	\end{center}
\end{figure}

In previous sections, we analyze the bifurcations that led to the emergence of new islands. Now, we describe the destruction of the islands, a transition related to the boundary between the colored and the black regions on the parameter spaces of Figure \ref{fig2}. From our analysis, we are able to conclude that all the islands pass through a pitchfork bifurcation and become islands of period two, with their centers at an equidistant distance from $y=0$. The destruction route is shown in Figure \ref{fig6} for $K_1=0.1$, $m_1=2$, $m_2=3$ and three different values of $K_2$.
	
In Figure \ref{fig6} (a), for $K_2=1.5$, we observe the three islands related to the prevalent mode $m_2=3$. Visually, the island in the center is different from the other two. As the value of $K_2$ increases, all three islands become elongated in the $y$-direction. If $K_2$ continues to increase, the width of the island in the $x$-direction shrinks and a pitchfork bifurcation occurs, where the stability of the elliptic point changes and a hyperbolic point emerges. We observe the bifurcation at the phase space in Figure \ref{fig6} (b). For $K_2=4.0$, the two islands with the center in $x\approx\pm0.35$ are elongated in the $y$-direction, but their centers are elliptic points. In contrast, the elliptic point centered in $x=0$ bifurcates, one hyperbolic point emerges in its place, along with two new elliptic points, as we can observe in the amplification of around it in the inset. For greater values of $K_2$, the two reminiscent elliptic points also bifurcate, and the final state is shown in Figure \ref{fig6} (c), where instead of three islands with period one with the center in $y=0$, we observe three islands of period two with their centers at an equidistant distance from $y=0$. Therefore, in the parameter spaces in Figure \ref{fig2}, the black points do not indicate necessarily the non-existence of islands in the phase space, but rather that there are no elliptic points in $y=0$.

 \subsection{Transition routes}
Observing the destruction route in Figure \ref{fig6}, we conclude that different islands can disappear for different parameter values. Just as observed in Figure \ref{fig2} (b) where there is an intermediate mode, there may be one or more intermediate modes in the destruction routes. In Tables 1-3, we summarize all the observed transitions for every combination of $m_1\in[1,5]$ and $m_2 \in[m_1+1,6]$ in the three routes from mode $m_1$ to $m_2$ and for the destruction routes from $m_2$ to no elliptic points at $y=0$. We remember that the decrease in the number of elliptic points always happens through the pitchfork bifurcation.
	\begin{table}[]
		\begin{minipage}{0.45\textwidth}
			\centering
			\begin{tabular}{c|c|l}
				\hline
				\addlinespace
				\multicolumn{3}{c}{Route 1: $m_1 \xrightarrow{\text{P}} m_2$}\\
				\hline
				$m_1$& $m_2$ & Transitions\\
				\hline
				1 & 2& $1 \to 2 \to 1 \to 0$\\
				2 & 3 & $2 \to 3 \to 2 \to 0$ \\
				2 & 4 &$ 2 \to 4 \to 3 \to 2 \to 0$\\
				3 & 4 & $3 \to 4 \to 3 \to 1 \to 0$\\	
				3 & 6 & $3 \to 6 \to 3 \to 0$\\
				4 & 5 & $4 \to 5 \to 4 \to 2 \to 0$\\
				4 & 6 & $4 \to 6 \to 4 \to 0$\\
				5 & 6 & $5 \to 6 \to 5 \to 3 \to 1 \to 0$\\
				\hline
			\end{tabular}
			\caption{Transitions in third column for each combination of $m_1$ and $m_2$ in which the route that leads from one mode to the other is only one pitchfork bifurcation. The letter P stands for pitchfork bifurcation.}
			\label{tab1}
		\end{minipage}
		\hspace{0.05\textwidth}
		\begin{minipage}{0.45\textwidth}
			\centering
			\begin{tabular}{c|c|l}
				\hline
				\addlinespace
				\multicolumn{3}{c}{Route 2: $m_1 \xrightarrow{\text{SN}} m_2$}\\
				\hline
				$m_1$& $m_2$ & Transitions\\
				\hline
				1 & 3 & $1 \to 3 \to 2 \to 0$ \\ 
				1 & 5& $1 \to 5 \to 4 \to 2 \to 0$ \\         
				2 & 6 & $2 \to 6 \to 4 \to 0$\\
				3 & 5 & $3 \to 5 \to 4 \to 2 \to 0$\\
				\hline
			\end{tabular}
			\caption{For the combinations of modes $m_1$ and $m_2$ shown in the first and second columns, we indicate the transitions that occur. In this case, Route 2, only on saddle-node (SN) bifurcation happens to take the system from mode $m_1$ to mode $m_2$.}
			\label{tab2}
		\end{minipage}
	\end{table}
	
	\begin{table}[]
		\centering
		\begin{tabular}{c|c|l}
			\hline
			\addlinespace
			\multicolumn{3}{c}{Route 3: $m_1 \xrightarrow{\text{P + SN}} m_2$}\\
			\hline
			$m_1$& $m_2$ & Transitions\\
			\hline
			1 & 4& $1 \to 2 \to 4 \to 3 \to 1 \to 0$\\ 
			1 & 6 &$ 1 \to 2 \to 4 \to 6 \to 5 \to 3  \to 2 \to 0$\\
			2 & 5 & $2 \to 3 \to 5 \to 4 \to 2 \to 0$\\
			\hline
		\end{tabular}
		\caption{In Route 3, we observe intermediate modes. Above, we present all the modes for the combinations of $m_1$ and $m_2$ shown in the first two columns. In this scenario, the route is composed of one pitchfork (P) bifurcation, and it is followed  by consequently saddle-nodes (SN) bifurcations until the system assumes the mode $m_2$.}
		\label{tab3}
	\end{table}
	
Table \ref{tab1} presents the transitions that occur when the transition from mode $m_1$ to $m_2$ occurs through Route 1, i.e., from only one pitchfork bifurcation. In Table \ref{tab2}, we have the transition related to Route 2. The transition from mode $m_1$ to $m_2$ occurs by the saddle-node bifurcation without intermediate mode. Lastly, in Table \ref{tab3}, we present the transition related to Route 3, where intermediate modes are identified. 

The destruction of elliptic points in $y=0$ always occurs by pitchfork bifurcations. Observing all the transitions in Tables \ref{tab1}-\ref{tab3} from mode $m_2$ to zero elliptic points in $y=0$, we can not identify a general pattern or specific sequence. We intend to investigate deeply the breakup of these islands in future research.

\newpage 
\section{Conclusions}
To propose a simple system that exhibits isochronous bifurcations, we define the two-harmonic standard map, a two-dimensional map with two parameters where two different arbitrary resonant modes compete. From the phase space analysis, we identified the emergence of new islands as the amplitude of the second mode increases. The new islands can appear within or outside the original set of islands.
	
From our observations of the phase space, the emergence of new islands and the transition from the first mode $m_1$ to the second mode $m_2$ can happen in three ways. The first way, that we named Route 1, is through pitchfork bifurcation. In this case, several fixed points change their stability and twice as many fixed points emerge with the opposite stability. In the example presented, one elliptic point becomes hyperbolic and two new elliptic points emerge. Route 2 is similar to Route 1, with the difference that, in this case, the bifurcation is a saddle-node bifurcation. In this scenario, as the amplitude of the second mode increases,  hyperbolic and elliptic points are created in the bifurcation point. Lastly, we identified Route 3 where intermediate modes exist. For the last route, we observe one pitchfork bifurcation and then sequential saddle-node bifurcation until the system reaches the second mode. With this analysis, we can affirm that the two harmonic standard maps can simulate different types of isochronous bifurcations.

Differently for the emergence of islands, our observations showed that the destruction of islands always occurs through the same bifurcation type, the pitchfork bifurcation. For the destruction, the width of the island in the $x$-direction shrinks while the length in the  $y$-direction increases. Then, the elliptic point goes through a pitchfork bifurcation, becomes hyperbolic, and two islands emerge, leading to a scenario where there are no elliptic points in $y=0$.
	
	

\section{Acknowledgments} 

This research recieved the support of the Coordination for the Improvement of Higher Education Personnel (CAPES) under Grant No. 88887.320059/2019-00, 88881.143103/2017-01, the National Council for Scientific and Technological Development (CNPq - Grant No. 403120/2021-7, 311168/2020-5, 301019/2019-3) and Fundação de Amparo à Pesquisa do Estado de São Paulo (FAPESP) under Grant No. 2022/12736-0, 2018/03211-6, 2023/10521-0. We would also like to thank the 105 Group Science \cite{105GS} for fruitful discussions.

\section{Author statement}


\subsubsection*{Declaration of interests}

The authors declare that they have no known competing financial interests or personal relationships that could have appeared to influence the work reported in this paper.

\subsubsection*{Declaration of generative AI in scientific writing}
Generative AI/AI assisted technologies were not used in the writing process of this manuscript.

\subsubsection*{CRediT author statement}
\textbf{Michele Mugnaine:} Conceptualization, Formal analysis, Investigation, Methodology, Validation, Visualization, Writing - original draft, review and editing. \textbf{Bruno B. Leal:} Conceptualization, Formal analysis, Investigation, Methodology, Validation, Visualization, Writing - original draft, review and editing. \textbf{Alfredo M. Ozorio de Almeida:} Methodology, Validation, Visualization, Writing - original draft, review and editing. \textbf{Ricardo L. Viana:} Methodology, Validation, Visualization, Writing - original draft, review and editing. \textbf{Iberê L. Caldas:} Conceptualization, Formal analysis, Investigation, Methodology, Validation, Visualization, Writing - original draft, review and editing.

\subsubsection*{Data availability}
The source code and data are openly available online in the Oscillations Control Group Data
Repository \cite{OCG}.

\textbf{}
	\bibliographystyle{ieeetr}
	\bibliography{refs}
\end{document}